\documentclass[aps,prd,twocolumn,showpacs,amsmath,amssymb,nofootinbib,eqsecnum, preprintnumbers]{revtex4-1}
\usepackage{graphicx}
\usepackage{epstopdf}

\newcommand{\be}{\begin{equation}}
\newcommand{\ee}{\end{equation}}
\newcommand{\ba}{\begin{eqnarray}}
\newcommand{\ea}{\end{eqnarray}}

\newcommand{\beq}{\begin{equation}}
\newcommand{\eeq}{\end{equation}}
\newcommand{\beqa}{\begin{eqnarray}}
\newcommand{\eeqa}{\end{eqnarray}}

\begin{document}
\title{$P-V$ criticality of charged AdS black holes}

\author{David Kubiz\v n\'ak}
\email{dkubiznak@perimeterinstitute.ca}
\affiliation{Perimeter Institute, 31 Caroline St. N. Waterloo
Ontario, N2L 2Y5, Canada}

\author{Robert B. Mann}
\email{rbmann@sciborg.uwaterloo.ca}
\affiliation{Department of Physics and Astronomy, University of Waterloo,
Waterloo, Ontario, Canada, N2L 3G1}

\date{May 1, 2012}  

\begin{abstract}
Treating the cosmological constant as a thermodynamic pressure and its conjugate quantity as a thermodynamic volume, we reconsider the critical behaviour of charged AdS black holes. We complete the analogy of this system with the liquid--gas system and 
study its critical point, which occurs at the point of divergence of specific heat at constant pressure. We calculate the critical exponents and show that they coincide with those of the Van der Waals system.
\end{abstract}


\maketitle

\section{Introduction}
Thermodynamic properties of black holes have been studied for many years. During that time it turned out that black hole spacetimes can not only be assigned standard thermodynamic variables such as temperature or entropy, but were also shown to possess rich phase structures and admit critical phenomena, in complete analogy with known non-gravitational thermodynamic systems elsewhere in nature. 
Of especial  interest  is  black hole  thermodynamics in the presence of a negative cosmological constant. {\em Asymptotically AdS} black hole spacetimes admit a gauge duality description and are described by dual thermal field theory. Correspondingly one has a microscopic description of the underlying degrees of freedom at hand. 
This duality has been recently exploited to study  the behaviour of  quark-gluon plasmas  and  for the qualitative description of various condensed matter phenomena.  

The history of the study of thermodynamic properties of AdS black holes began with the seminal paper of Hawking and Page \cite{HawkingPage:1983}, who demonstrated the existence of a certain phase transition in the phase space of the (non-rotating uncharged) Schwarzschild-AdS black hole. Since then our understanding of phase transitions and critical phenomena has been extended to a variety of more complicated backgrounds, e.g., \cite{CveticGubser:1999a, CveticGubser:1999b}. Of particular  interest is the discovery of the first order phase transition in the charged (non-rotating) Reissner--Nordstr\"om-AdS (RN-AdS) black hole spacetime \cite{ChamblinEtal:1999a, ChamblinEtal:1999b}. This transition displays classical critical behaviour and is {superficially {\em analogous} to} a liquid--gas phase transition {\em \`ala} Van der Waals. We critically review this analogy in the appendix.

Recently, the idea of including the variation of the cosmological constant $\Lambda$ in the first law of black hole thermodynamics has attained increasing  
attention \cite{CaldarelliEtal:2000, KastorEtal:2009, Dolan:2010, Dolan:2011a, Dolan:2011b, CveticEtal:2011, LuEtal:2012}.\footnote{In fact, this idea was first applied (and is now commonly used)  in cosmology. For black hole physics it was seriously established in \cite{KastorEtal:2009}.}
From a general relativistic point of view, such a variation is a slightly awkward thing to do, as the cosmological constant should be treated as a fixed external parameter of the theory. Moreover, varying $\Lambda$ in the first law corresponds to comparing black hole ensembles with different asymptotics. This is different from standard thermodynamic considerations, where the black hole parameters  are varied in a `fixed AdS background'. However, there exist at least three good reasons why the variation of $\Lambda$ should be included in thermodynamic considerations: i) One may consider `more fundamental' theories, where physical constants, such as Yukawa couplings, gauge coupling constants, Newton's constant, or the cosmological constant are not fixed a priori, but arise as vacuum expectation values and hence can vary. In such a case it is natural to include variations of these `constants' in the first law, e.g., \cite{GibbonsEtal:1996, CreightonMann:1995}.  ii) More pragmatically, in the presence of a cosmological constant the first law of black hole thermodynamics becomes inconsistent with the Smarr relation (the scaling argument is no longer valid) unless the variation of $\Lambda$ is included in the first law \cite{KastorEtal:2009}. A similar situation occurs, for example, for Born--Infeld black holes, where the variation of the maximal field strength $b$ must be included in the first law for it to be consistent with the corresponding Smarr relation \cite{Rasheed:1997,Breton:2004, Huan:2010}. iii) Once the variation of $\Lambda$ is included in the first law, the black hole mass $M$ is identified with {\em enthalpy} rather than internal energy \cite{KastorEtal:2009}. It is also natural to consider a variable thermodynamically conjugate to $\Lambda$. 
Since $\Lambda$ corresponds to pressure,
its conjugate variable has dimensions of volume, characteristic for a given black hole spacetime.  
In particular, using   geometric units $G_N=\hbar=c=k=1$, in the case of an asymptotically AdS black hole in four dimensions, one identifies the pressure with
\be\label{P}
P=-\frac{1}{8\pi} \Lambda=\frac{3}{8\pi}\frac{1}{l^2}\,,
\ee
and the corresponding black hole `{\em thermodynamic volume}' with $V=\left(\frac{\partial M}{\partial P}\right)_{S, Q_i, J_k}$. In the simplest case of a Reissner-Nordstr\"om-AdS (RN-AdS) black hole the volume is given by  
\be\label{V}
V=\frac{4}{3}\pi r_+^3\,,
\ee
with $r_+$ being the radius of the black hole event horizon expressed in
terms of the `Schwarzschild radial coordinate'. More generally, the thermodynamic volume for a wide variety of known black hole solutions was studied in \cite{CveticEtal:2011}. It turns out that this volume seems to possess some universal properties 
and is  conjectured to satisfy the {\em reverse isoperimetric inequality}. If true, such an inequality may restrict some processes in black hole spacetimes.  

Taking these considerations seriously, once the  black hole  pressure and volume are identified, one may proceed to calculate various thermodynamic quantities employing   standard thermodynamic machinery. For example, one may study adiabatic compressibility, specific heat at constant pressure, or even the `speed of sound' associated with the black hole \cite{Dolan:2010, Dolan:2011a, Dolan:2011b}. Moreover, as noted by Dolan \cite{Dolan:2011a}, this also opens an interesting possibility of reconsidering the critical behaviour of AdS black holes in an {\em extended phase space}, including pressure and volume as thermodynamic variables. In particular, Dolan studied $P=P(V,T)$ equation of state for a rotating charged AdS black hole, observed an analogy with the Van der Waals $P-V$ diagram, and determined its critical point. 

In this paper we explore this issue further and elaborate on Dolan's results. For simplicity, we concentrate on studying $P-V$ criticality of (spherical) RN-AdS black holes. Our aim is to complete the identification  
with the liquid--gas system in the extended phase space, and to clarify the relationship with the previously observed Van der Waals-like behaviour of the  system discussed in  \cite{ChamblinEtal:1999a, ChamblinEtal:1999b}. For this purpose, we study the behaviour of the Gibbs free energy in the {\em fixed charge} ensemble. Equipped with this we obtain the phase transition in the $(P, T)$-plane, standard for the liquid--gas transition.  {While 
this behaviour is very similar to that uncovered in \cite{ChamblinEtal:1999a, ChamblinEtal:1999b}, the results we obtain have
a rather different interpretation -- they are not an {\em analogy}, but actually compare the `same physical quantities'.} For {example we find, as in the liquid--gas system, that the} transition occurs in the $(P,T)$-plane and not in the $(\beta, Q)$-plane.
We also study the critical exponents and show that they coincide with those of the Van der Waals system. This completes the identification of the charged black hole first order transition with the standard liquid-gas phase transition. 

Our paper is organized as follows. In the next section we review some basic facts about the Van der Waals fluid and its critical exponents. {In Sec. 3 we demonstrate analogous behaviour of the RN-AdS black hole system to the Van der Waals fluid. Specifically,
 $P-V$ criticality}, Gibbs free energy, first order phase transition, and the behaviour near the critical point are identified with the liquid--gas system.
Sec. 4 is devoted to conclusions. To compare our results with the Van der Waals analogy studied in \cite{ChamblinEtal:1999a, ChamblinEtal:1999b} we include an appendix.

\section{Van der Waals fluid}
Van der Waals equation, e.g. \cite{Goldenfeld:1992},  is a popular closed form modification of the ideal gas law, which approximates the behaviour of real fluids, taking into account the nonzero size of molecules 
and the attraction between them. It is often used to describe basic qualitative features of the liquid--gas phase transition. 
The equation reads
\be\label{vdwA}
\left(P+\frac{a}{v^2}\right)(v-b)=kT\,.
\ee
Here, $v=V/N$ is the specific volume of the fluid, $P$ its pressure, $T$ its temperature, and $k$ is the Boltzmann constant. 
The constant $b>0$ takes into account the nonzero size of the molecules of a given fluid, whereas the constant $a>0$ is a measure of the attraction  between them. 
Equivalently, one can expand this equation to write it as a cubic equation for $v$, 
\be\label{vdwA2}
Pv^3-(kT+bP)v^2+av-ab=0\,.
\ee
\begin{figure}\label{fig:Fig1}
\begin{center}
\rotatebox{-90}{
\includegraphics[width=0.39\textwidth,height=0.34\textheight]{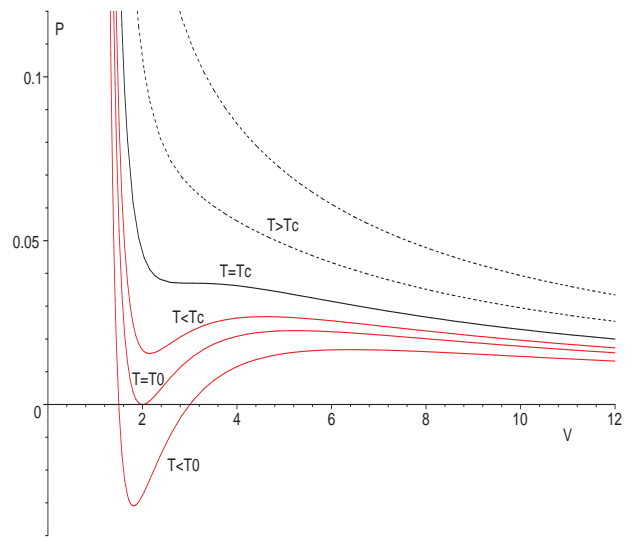}
}
\caption{{\bf $P-V$ diagram of Van der Waals fluid.} The temperature of isotherms decreases from top to bottom. The two upper dashed lines correspond to the ``ideal gas'' phase for $T>T_c$, the critical isotherm $T=T_c$ is denoted by the thick solid line, lower solid lines correspond to temperatures smaller than the critical temperature; $T=T_0$ isotherm is also displayed. The constants $a$ and $b$ 
in Eq. (\ref{vdwA})
were set equal to one.  
}  
\end{center}
\label{pVVdW}
\end{figure} 

The qualitative behaviour of isotherms in the $P-V$ diagram is depicted in Fig.~1. The {\em critical point} occurs when $P=P(v)$ has an inflection point, i.e.,  
\be
\frac{\partial P}{\partial v}=0\,,\quad \frac{\partial^2 P}{\partial v}=0\,,
\ee
at the critical isotherm $T=T_c$.\footnote{Note that for $T<T_0=\frac{1}{4k}\frac{a}{b}=\frac{27}{32} T_c$, a part of the 
isotherm corresponds to a negative pressure. Similar to the black hole case discussed in the next section, this is unphysical and once the Maxwell's equal area law, \eqref{Maxwell}, is employed this is replaced by the corresponding isobar of positive pressure.
} 
Writing \eqref{vdwA2} in the form $P_c(v-v_c)^3=0$\,, and comparing the coefficients
we conclude that  
\be\label{abc}
kT_c=\frac{8a}{27b}\,,\quad v_c=3b\,,\quad P_c=\frac{a}{27b^2}\,.
\ee
We also note that
\be\label{univ}
\frac{P_c v_c}{k T_c}=\frac{3}{8}
\ee
is a universal number predicted for all fluids (independent of constants $a$ and $b$).
Defining
\be
p=\frac{P}{P_c}\,,\quad \nu=\frac{v}{v_c}\,,\quad \tau =\frac{T}{T_c}\,,
\ee
we obtain the  so called  {\em law of corresponding states} 
\be\label{states}
8\tau=(3\nu-1)\left(p+\frac{3}{\nu^2}\right)\,.
\ee
This law is universal and is valid under more general assumptions than the original derivation of the Van der Waals equation.

For $T<T_c$ there is a {\em liquid--gas phase transition} in the system. To describe this phase transition one has to replace the `oscillating part' of the isotherm by an isobar, according to  {\em Maxwell's equal area law}
\begin{figure}
\begin{center}
\rotatebox{-90}{
\includegraphics[width=0.39\textwidth,height=0.34\textheight]{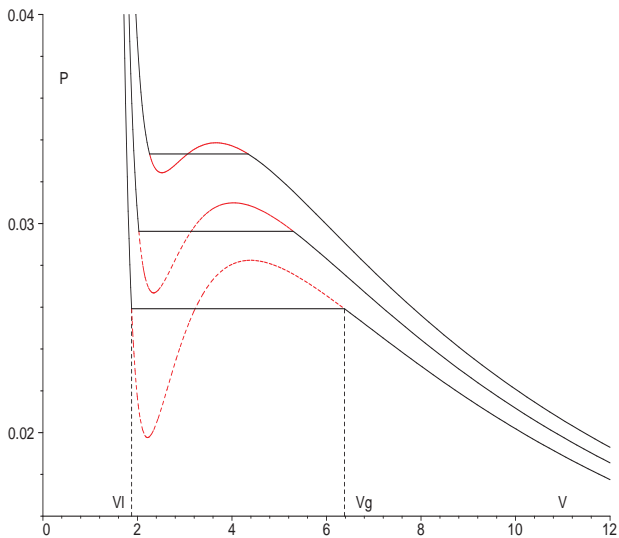}
}
\caption{{\bf Maxwell's equal area law.} The `oscillating' (dashed) part of the isotherm $T<T_c$ is replaced by an isobar, such that the areas above and below the isobar are equal one another.  
}  
\end{center}
\label{MaxwellVdW}
\end{figure} 
\be\label{Maxwell}
\oint v dP=0\,,
\ee
as shown in Fig.~2.
This prescription reflects the fact that at the transition both phases 
have the same Gibbs free energy, see \eqref{dG}. Eq. \eqref{Maxwell} is an effective tool to find the pressure at which the phase transition occurs, or to calculate the change of the volume and the latent heat.

To get more information about the phase transition we study the (specific) Gibbs free energy, $G=G(P,T)$. 
For a fixed number of particles, this can be obtained by integrating its differential  
\be\label{dG}
dG=-SdT+v dP\,,
\ee
while using the Van der Waals equation. The unspecified integration function can be determined from comparing with (statistical) $G$ of the ideal gas.   
The recovered specific Gibbs free energy reads 
\be
G=G(T,P)=-kT\left(1+\ln\left[\frac{(v-b)T^{3/2}}{\Phi}\right]\right)-\frac{a}{v}+Pv\,.
\ee
Here, $v$ is understood as a function of $P$ and $T$, through the Van der Waals equation \eqref{vdwA}, and $\Phi$ is a (dimensionful) constant characterizing the gas.  The Gibbs free energy is depicted in Fig.~3 and Fig.~4.
\begin{figure}
\begin{center}
\rotatebox{-90}{
\includegraphics[width=0.39\textwidth,height=0.30\textheight]{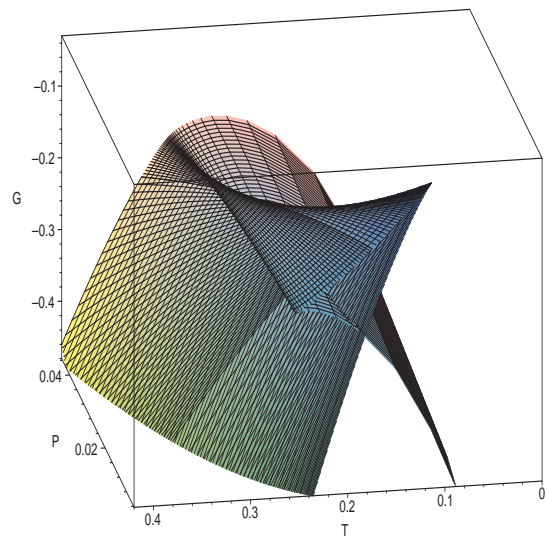}
}
\caption{{\bf Gibbs free energy of Van der Waals fluid.}  This picture shows  the characteristic swallowtail behaviour of the Gibbs free energy as a function of pressure and temperature. This corresponds to a first-order liquid--gas phase transition which occurs at the intersection of $G$ surfaces. The corresponding curve is called the coexistence line. We have set $\Phi=1$.
}  
\end{center}
\end{figure} 
\begin{figure}
\begin{center}
\rotatebox{-90}{
\includegraphics[width=0.39\textwidth,height=0.34\textheight]{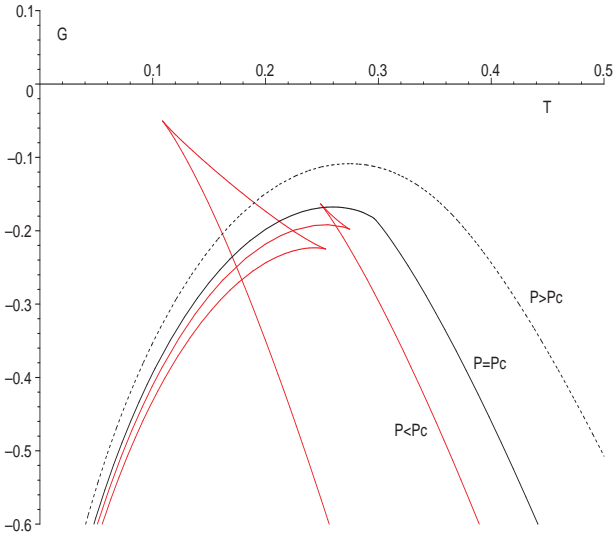}
}
\caption{{\bf Gibbs free energy of Van der Waals fluid.} 
This figure depicts the qualitative behaviour of the Gibbs free energy as a function of temperature for various pressures. The pressure decreases from right to left. The dashed line corresponds to $P=2P_c$, the thick solid line to $P=P_c$, and the remaining solid lines display $P=0.6 P_c$ and $P=0.09 P_c$, respectively. For $P<P_c$ there is a first-order transition in the system.
}  
\end{center}
\end{figure} 

The {\em coexistence line} of two phases, along which these two phases are in equilibrium, occurs where two surfaces of $G$ cross each other. 
This line is depicted in Fig.~5. It is governed by the Clausius--Clapeyron equation 
\be\label{CC}
\frac{dP}{dT}\Big|_{\mbox{coexistence}}=\frac{S_g-S_l}{v_g-v_l}\,,
\ee 
where $S_g$, $S_l$, and $v_g$, $v_l$, stand for the specific entropy and specific volume of the gas, liquid, phase, respectively. 
Alternatively, this line can be constructed by exploiting Maxwell's law \eqref{Maxwell}, or by finding a curve in the $(P,T)$-plane for which the Gibbs free energy and the Van der Waals temperature coincide for two different volumes $v_l$ and $v_g$. 
\begin{figure}
\begin{center}
\rotatebox{-90}{
\includegraphics[width=0.39\textwidth,height=0.34\textheight]{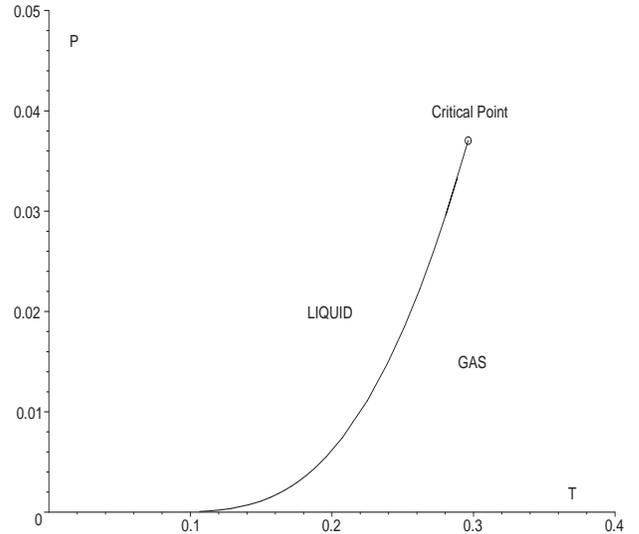}
}
\caption{{\bf Coexistence line of liquid--gas phase.} Fig. displays the coexistence line of liquid and gas phases of the Van der Waals fluid in $(P, T)$-plane. The critical point is highlighted by a small circle at the end of the coexistence curve.  
}  
\end{center}
\end{figure}

{\em Critical exponents} describe the behaviour of physical quantities near the critical point. It is believed that they are {\em universal}, i.e., they do not depend on the details of the physical system.\footnote{They may, however, depend on the dimension of the system or the range of the interactions. In $D\geq 4$ dimensions they can be calculated by using the mean field (Landau's) theory, e.g., \cite{Goldenfeld:1992}.
}
Let us calculate a few of these exponents as predicted by the equation of corresponding states.
We denote 
\be
t=\frac{T-T_c}{T_c}=\tau-1\,,\quad \phi=\frac{v-v_c}{v_c}=\nu-1\,. 
\ee
The critical exponents $\alpha, \beta, \gamma, \delta$ are defined as follows:
\begin{itemize}
\item
Exponent $\alpha$ governs the behaviour of the specific heat at constant volume, 
\be
C_v=T \frac{\partial S}{\partial T}\Big|_{v}\propto |t|^{-\alpha}\,.
\ee
\item Exponent $\beta$ describes the behaviour of the {\em order parameter} $\eta=v_g-v_l$ (the difference of the volume of the gas $v_g$ phase and the volume of the liquid phase $v_l$)  on the given isotherm
\be
\eta =v_g-v_l\propto |t|^\beta\,.
\ee
\item 
Exponent $\gamma$ determines the behaviour of the {\em isothermal compressibility} $\kappa_T$, 
\be
\kappa_T=-\frac{1}{v}\frac{\partial v}{\partial P}\Big|_T\propto |t|^{-\gamma}\,.
\ee
\item Exponent $\delta$ governs the following behaviour on the critical isotherm $T=T_c$:  
\be
|P-P_c|\propto |v-v_c|^\delta\,.
\ee
\end{itemize}

Let us now, following \cite{Goldenfeld:1992}, calculate these exponents for our system. 
To calculate $C_v$ we consider the specific free energy 
\be
F(T,v)=G-Pv=-kT\left(1+\ln\left[\frac{(v-b)T^{3/2}}{\Phi}\right]\right)-\frac{a}{v}\,.
\ee
From here we can calculate the specific entropy  
\be
S(T,v)=-\left(\frac{\partial F}{\partial T}\right)_{\!v}
=k\left(\frac{5}{2}+\ln\left[\frac{(v-b)T^{3/2}}{\Phi}\right]\right)\,,
\ee
and hence 
\be
C_v=T \frac{\partial s}{\partial T}\Big|_{v}=\frac{3k}{2}\,.
\ee
Since this expression is independent of $t$ we conclude that exponent $\alpha=0$.

Expanding the equation of corresponding states, \eqref{states}, in the vicinity of the critical point we have 
\be\label{expansion}
p=1+4t-6t\phi-\frac{3}{2}\phi^3+O(t\phi^2, \phi^4)\,.
\ee
[The fact that we can neglect the term $O(t\phi^2)$ in this expansion is justified by Eq. \eqref{betaExp} below.] 
Differentiating this for fixed $t<0$ we get 
\be
dP=-P_c\left(6t+\frac{9}{2}\phi^2\right)d\phi\,.
\ee
Let us denote  $\phi_g=({v_g-v_c})/{v_c}$ the `volume' of gas and similarly $\phi_l$ the volume of the liquid. Using 
Maxwell's area law \eqref{Maxwell}, and the fact that during the phase transition the pressure remains constant, we have the following two equations:
\ba
p&=&1+4t-6t\phi_l-\frac{3}{2}\phi_l^3=1+4t-6t\phi_g-\frac{3}{2}\phi_g^3\,,\nonumber\\
0&=&\int_{\phi_l}^{\phi_g}\phi(6t+\frac{9}{2}\phi^2)d\phi\,.
\ea
These equations have a unique non-trivial solution, given by $\phi_g=-\phi_l=2\sqrt{-t}$. Hence we have 
\be\label{betaExp}
\eta=v_c(\phi_g-\phi_l)=2v_c\phi_g=4v_c\sqrt{-t}\quad \Rightarrow \quad \beta=1/2\,.
\ee

To calculate exponent $\gamma$, we differentiate Eq. \eqref{expansion} to get 
\be
\frac{\partial v}{\partial P}\Big|_T=-\frac{v_c}{6P_c}\frac{1}{t}+O(\phi)\,.
\ee
Hence we have 
\be
\kappa_T=-\frac{1}{v}\frac{\partial v}{\partial P}\Big|_T\propto \frac{1}{6P_c}\frac{1}{t} \quad \Rightarrow \quad 
\gamma=1\,.
\ee
We also note that using the formula 
\be\label{CpCv}
C_P-C_v=-T\left(\frac{\partial P}{\partial T}\right)^2_v \left(\frac{\partial v}{\partial P}\right)_T \propto \frac{k}{t}\,.
\ee
That is, $C_p$ diverges at the critical point with the same critical exponent $\gamma$ as $\kappa_T$.

Finally, the `shape' of the critical isotherm $t=0$ is given by \eqref{expansion}\,,i.e.,
\be
p-1=-\frac{3}{2}\phi^3\quad \Rightarrow \quad \delta=3\,.
\ee 
This completes the calculation of basic critical exponents for the Van der Waals fluid.

\section{Phase transition in charged AdS black hole spacetime} 
\subsection{Charged AdS black hole} 

To start with we review some basic thermodynamic properties of the spherical RN-AdS black hole. In Schwarzschild-like coordinates the metric and the $U(1)$ field  read  
\ba
ds^2&=&-Vdt^2+\frac{dr^2}{V}+r^2d\Omega_2^2\,,\\
F&=&dA\,,\quad A=-\frac{Q}{r} dt\,.
\ea 
Here, $d\Omega^2_2$ stands for the standard element on $S^2$ and function $V$ is given by
\be
\quad V=1-\frac{2M}{r}+\frac{Q^2}{r^2}+\frac{r^2}{l^2}\,.
\ee
With this choice we have a solution of the Einstein--Maxwell system of equations following from the bulk action
\be\label{IEM}
I_{EM}=-\frac{1}{16\pi}\int_M \sqrt{-g}\left(R-F^2+\frac{6}{l^2}\right)\,.
\ee

The position of the black hole event horizon is determined as a larger root of $V(r_+)=0$. 
The parameter $M$ represents the ADM mass of the black hole and in our set up it is associated with the enthalpy of the system.
$Q$ represents the total charge. 
Using the `Euclidean trick', one can identify the black hole temperature 
\be\label{T}
T=\frac{1}{\beta}=\frac{V'(r_+)}{4\pi}=\frac{1}{4\pi r_+}\left(1+\frac{3r_+^2}{l^2}-\frac{Q^2}{r_+^2}\right)\,,
\ee
and the corresponding entropy
\be\label{S}
S=\frac{A}{4}\,,\quad A=4\pi r_+^2\,.
\ee
The electric potential $\Phi$, measured at infinity with respect to the horizon, is 
\be
\Phi=\frac{Q}{r_+}\,.
\ee 
The thermodynamic volume $V$ of the black hole is given by \eqref{V} and the corresponding pressure $P$ by \eqref{P}. 
With these relations, the solution obeys the first law of black hole thermodynamics in an {\em extended} (including $P$ and $V$ variables)  {\em phase space}
\be\label{dM}
dM=TdS+\Phi dQ+VdP\,.
\ee
The corresponding Smarr relation 
\be
M=2(TS-VP)+\Phi Q
\ee
can be derived from it by a scaling (dimensional) argument  \cite{KastorEtal:2009}. 
When $P$ is treated as constant (i.e. if the cosmological constant is not allowed to vary), \eqref{dM} reduces to the standard first law in the `non-extended' phase space. In this case, however, the Smarr relation remains unchanged and no longer follows from the first law by the scaling argument. 

Previous work on the critical behaviour of RN-AdS black hole in the non-extended phase space demonstrated that in the canonical (fixed charge) ensemble, for $Q<Q_{c}$ and $\beta_Z<\beta<\beta_c$, there exists a first order phase transition in the system \cite{ChamblinEtal:1999a, ChamblinEtal:1999b}. This occurs in 
the $(\beta, Q)$-plane and the phase transition is quite analogous to the liquid--gas phase transition occurring in the $(P, T)$-plane for $T<T_c$.  We review this analogy in the appendix and find that it is not fully consistent. 

In what follows we concentrate on analyzing the phase transition of the AdS charged black hole system in the extended phase space
while we treat the black hole charge $Q$ as a fixed external parameter, not a  thermodynamic variable.   
We shall find that an even more remarkable coincidence with the Van der Waals fluid is realized in this case.

\subsection{Equation of state}
For a fixed charge $Q$, Eq. \eqref{T} translates into the 
{\em equation of state} for a charged AdS black hole,  $P=P(V,T)$,
\be\label{state}
P=\frac{T}{2r_+}-\frac{1}{8\pi r_+^2}+\frac{Q^2}{8\pi r_+^4}\,,\quad r_+=\left(\frac{3{V}}{4\pi}\right)^{1/3}\,.
\ee
Here, $V$ is the thermodynamic volume, given in terms of the event horizon radius $r_+$, $T$ is the black hole temperature, and $Q$ its charge.

 Before we proceed further we perform   dimensional analysis, translating the ``geometric'' equation of state  \eqref{state}  to a physical one. The physical pressure and temperature are given by 
\be
\mbox{Press}=\frac{\hbar c}{l_P^2} P\,,\quad \mbox{Temp}=\frac{\hbar c}{k} T\,,
\ee   
while the Planck length reads $l_P^2=\frac{\hbar G_N}{c^3}$. Multiplying \eqref{state} with $\frac{\hbar c}{l_P^2}$ we get
\be
\mbox{Press}=\frac{\hbar c}{l_P^2}P=\frac{\hbar c}{l_P^2}\frac{T}{2r_+}+\dots
=\frac{k\mbox{Temp}}{2l_P^2r_+}+\dots
\ee
Comparing with the Van der Waals equation, \eqref{vdwA}, we conclude that we should identify the   specific volume $v$ with 
\be
v=2l_P^2r_+\,.
\ee
In other words it is the horizon radius $r_+$, rather than the thermodynamic volume $V$, that should be associated with the fluid volume.


Pursuing this identification further and returning to geometric units, the equation of state \eqref{state} can now be written as\footnote{%
It is obvious that the criticality cannot happen in the `grand canonical' (fixed $\Phi$)  ensemble. Indeed, in that case the equation of state reduces to the quadratic equation
\be
Pv^2-8\pi Tv+4(1-\Phi^2)=0\,,
\ee
which does not admit critical behaviour. For a detailed discussion of thermodynamics in fixed $\Phi$ ensemble see \cite{PecaLemos:1999}. 
}
\be
P=\frac{T}{v}-\frac{1}{2\pi v^2}+\frac{2Q^2}{\pi v^4}\,,
\ee
The corresponding ``$P-V$ diagram'' is depicted in Fig~6. Evidently, for $Q\neq 0$  and for $T<T_c$ there is an inflection point and the behaviour is reminiscent  of the Van der Waals gas.
The {\em critical point} is obtained from 
\be
\frac{\partial P}{\partial v}=0\,,\quad \frac{\partial^2 P}{\partial v^2}=0\,,
\ee
which leads to 
\be\label{criticalvaluesAdS}
T_c=\frac{\sqrt{6}}{18\pi Q}\,,\quad v_c=2\sqrt{6} Q\,,\quad P_c=\frac{1}{96\pi Q^2}\,.
\ee
The critical radius corresponds to the critical thermodynamic volume 
\be
V_c=\frac{3}{4}\pi r_c^3=8\sqrt{6}\pi Q^3\,,
\ee 
which will be used later for calculating the critical exponents.
\begin{figure}
\begin{center}
\rotatebox{-90}{
\includegraphics[width=0.39\textwidth,height=0.34\textheight]{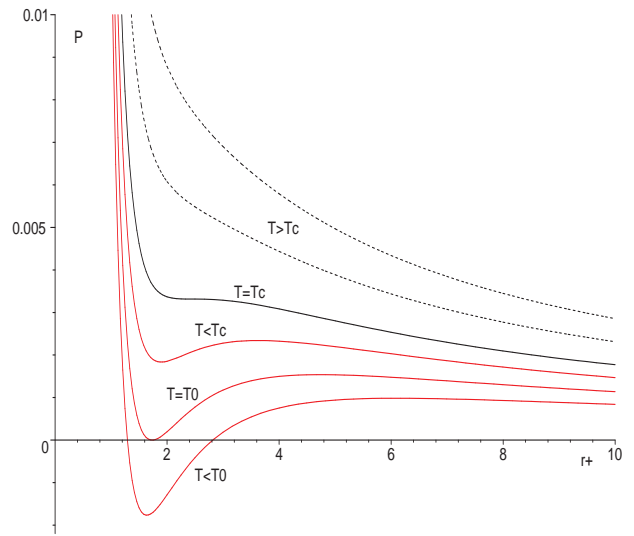}
}
\caption{{\bf $P-V$ diagram of   charged AdS black holes.}
The temperature of isotherms decreases from top to bottom. The two upper dashed lines correspond to the ``ideal gas'' one-phase behaviour for $T>T_c$, the critical isotherm $T=T_c$ is denoted by the thick solid line, lower solid lines correspond to temperatures smaller than the critical temperature, and the $T=T_0$ isotherm is also displayed. We have set $Q=1$.  
}  
\end{center}
\end{figure} 

Inspecting the critical values \eqref{criticalvaluesAdS} we find an interesting relation 
\be
\frac{P_c v_c}{T_c}=\frac{3}{8}\,,
\ee
which is exactly the same as for the Van der Waals fluid and is a universal number predicted for any RN-AdS black hole with arbitrary charge.   
Moreover  the  critical values \eqref{criticalvaluesAdS} can be written in the form \eqref{abc}, with  
\be
a=\frac{3}{4\pi}\,,\quad b=\frac{2\sqrt{6}Q}{3}\,.
\ee
It is the presence of the charge $Q$ that makes the effective volume smaller, $b\propto Q$.
Defining further
\be
p=\frac{P}{P_c}\,,\quad \nu=\frac{v}{v_c}\,,\quad \tau =\frac{T}{T_c}\,,
\ee
equation of state \eqref{state} translates into `the law of corresponding states' given by, cf. Eq. \eqref{states}, 
\be\label{statesR}
8\tau=3\nu\left(p+\frac{2}{\nu^2}\right)-\frac{1}{\nu^3}\,.
\ee

Similar to the Van der Waals equation (see Footnote~2), there exists a temperature $T_0$, given by
\be
T_0=\frac{\sqrt{3}}{18\pi Q}\,,
\ee
below which the pressure $P$ becomes negative for some $r_+$. As with the fluid this behaviour is unphysical, as one has to replace the oscillatory part of the 
isotherm by an isobar, according to   Maxwell's area law \eqref{Maxwell}. This is supported by studying the Gibbs free energy of the system in the next subsection.

Finally, let us mention that the critical behaviour is present only for a black hole  with horizon of spherical topology. 
For different topologies  the equation of state becomes 
\be\label{stateK}
P=\frac{T}{2r_+}-\frac{\hat{k}}{8\pi r_+^2}+\frac{Q^2}{8\pi r_+^4}\,,  
\ee
where  $\hat{k}=0$ is the toroidal (planar) case and $\hat{k}=-1$ is the higher-genus (hyperbolic) case.
Obviously, since all the terms on the r.h.s. are now non-negative,   critical behaviour cannot occur. Consequently there is also no temperature $T_0$.

\subsection{Gibbs free energy}
\begin{figure}
\begin{center}
\rotatebox{-90}{
\includegraphics[width=0.45\textwidth,height=0.34\textheight]{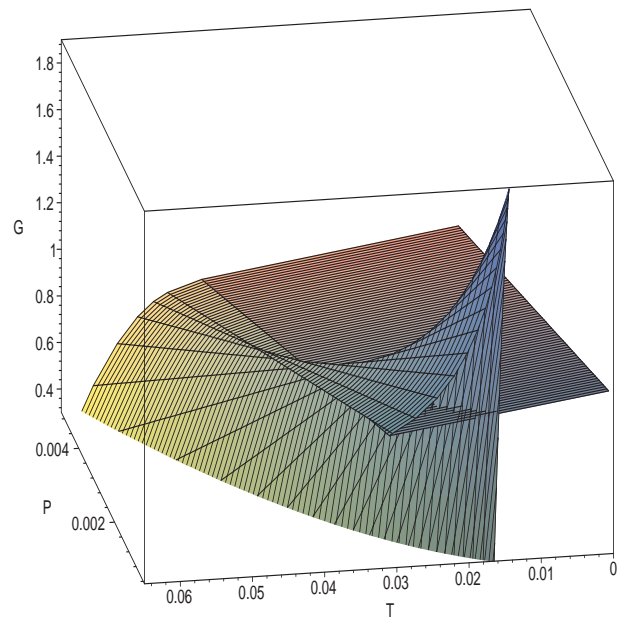}
}
\caption{{\bf Gibbs free energy of charged AdS black holes.} 
Figure shows characteristic swallowtail behaviour of the Gibbs free energy as a function of pressure and temperature for a fixed $Q=1$. The first-order small--large black hole transition occurs at a curve which is an intersection of $G$ surfaces.
}  
\end{center}
\end{figure} 
\begin{figure}
\begin{center}
\rotatebox{-90}{
\includegraphics[width=0.39\textwidth,height=0.34\textheight]{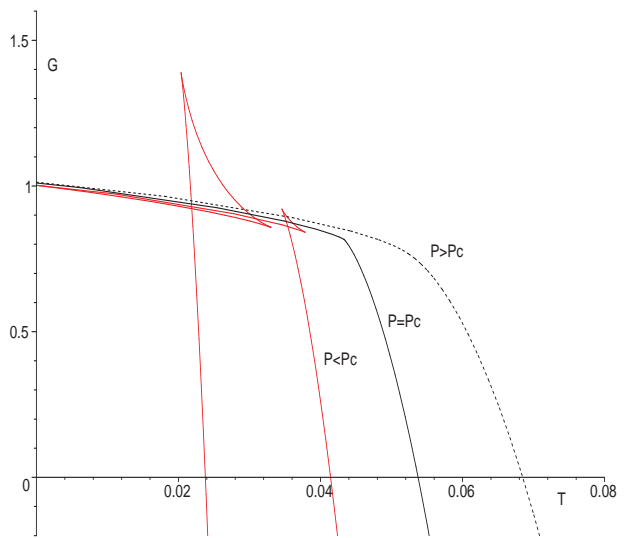}
}
\caption{{\bf Gibbs free energy of charged AdS black hole.} Fig. depicts $G$ as a function of temperature for fixed $Q=1$, and increasing pressure $P/P_c=0.1$ (black line) top to $P/P_c=1.6 $ (red line). The green line corresponds to critical pressure $P=P_c\approx 0.0033$. Obviously, for $T<T_c\approx 0.043$ there is a (small black hole)-(large black hole) first-order phase transition. 
}  
\end{center}
\end{figure} 
In order to find the partition function of the system let us calculate its Euclidean action. For a fixed charge $Q$, one considers a surface integral
\be
I_s=-\frac{1}{8\pi}\int_{\partial M} d^3x\sqrt{h}K-\frac{1}{4\pi}\int_{\partial M} d^3x \sqrt{h}n_a F^{ab} A_b\,.
\ee
The first term is the standard Gibbons--Hawking term while the latter term is needed to impose fixed $Q$ as a boundary condition at infinity.
The total action is then given by 
\be
I=I_{EM}+I_s+I_c\,,
\ee
where $I_{EM}$ is given by \eqref{IEM}, and $I_c$ represents the invariant counterterms needed to cure the infrared divergences \cite{EmparanEtal:1999, Mann:1999}.
 The total action was first calculated in  \cite{ChamblinEtal:1999b, CaldarelliEtal:2000} and reads  
 \be
I=\frac{\beta}{4l^2}\left(l^2r_+-r_+^3+\frac{3l^2Q^2}{r_+}\right)\,.
\ee
In the standard approach this action is associated with the free energy, stressing its dependence on extensive quantity $Q$. However, since we are considering an extended phase space, and the action has been calculated for   fixed $\Lambda$, following \cite{Dolan:2011a} we associate it with the {\em Gibbs free energy} for  fixed charge,  
\be\label{GibbsAdS}
G=G(T,P)=\frac{1}{4}\left(r_+-\frac{8\pi}{3} P r_+^3+\frac{3Q^2}{r_+}\right)\,.
\ee
Here, $r_+$ is understood as a function of pressure and temperature, $r_+=r_+(P,T)$, via equation of state \eqref{state}.

The behaviour of $G$ is depicted in Fig.~7 and Fig.~8. Since the $G$ surface demonstrates the characteristic `swallow tail' behaviour, there is a first order transition in the system.  The coexistence line in the $(P, T)$-plane can be obtained from  Maxwell's equal area law, the Clausius--Clapeyron equation \eqref{CC}, or simply by finding a  curve in the $(P,T)$-plane for which the Gibbs free energy and temperature coincide for small (with $r_+=r_s$) and large ($r_+=r_l$) black holes. This is depicted in Fig.~9.     
\begin{figure}
\begin{center}
\rotatebox{-90}{
\includegraphics[width=0.39\textwidth,height=0.34\textheight]{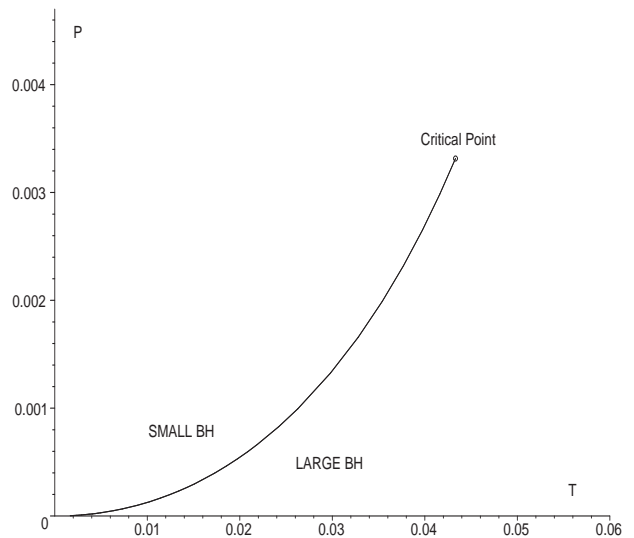}
}
\caption{{\bf Coexistence line of charged AdS black hole.} Fig. displays the coexistence line of small--large black hole phase transition of the charged AdS black hole system in $(P, T)$-plane. The critical point is highlighted by a small circle at the end of the coexistence line.
}  
\end{center}
\end{figure}

We note that the coexistence line in the $(P,T)$-plane and the one of the Van der Waals fluid in the previous section look  very much alike. The small--large black hole transition occurs for $T<T_c$, for any non-trivial value of the charge $Q$. Furthermore we
note some important differences between this diagram and an analogous diagram displayed in the appendix. There the transition occurs in $(Q, \beta)$-plane, for $Q<Q_c$ and $\beta<\beta_c$. Moreover there is a lower bound, Zorro's temperature $\beta_Z$, below which the transition no longer occurs. This analogy also `confuses' extensive and intensive thermodynamic quantities---see the appendix for more details.

\subsection{Behavior near critical point}
Let us now compute the critical exponents $\alpha, \beta, \gamma, \delta$ for the black hole system, following the discussion in the previous section.
We start with the behaviour of the specific heat at constant thermodynamic volume $C_V$. For this we consider the free energy 
\be
F(T,V)=G-PV=\frac{1}{2}\left(r_+-2\pi T r_+^2+\frac{Q^2}{r_+}\right)\,.
\ee
From here we calculate the entropy 
\be
S(T,V)=-\left(\frac{\partial F}{\partial T}\right)_V=\pi r_+^2\,,
\ee 
which coincides with \eqref{S}. Since this is independent of $T$ we find that $C_V=0$ and hence exponent $\alpha=0$.

Expanding around a critical point\footnote{Note that we could also write $\omega^\prime=\frac{v}{v_c}-1$ and the same
results for the critical exponents would be obtained.} 
\be
t=\frac{T}{T_c}-1\,,\quad \omega=\frac{V}{V_c}-1\,,
\ee 
we approximate  the law of corresponding states \eqref{statesR}
as
\be\label{stateApV}
p=1+\frac{8}{3}t -\frac{8}{9}t\omega -\frac{4}{81}\omega^3+O(t\omega^2, \omega^4)\,.
\ee
[The truncation of the series is justified by formula \eqref{etaAdS} below.] 
Differentiating this series for a fixed $t<0$ we get   
\be
dP=-\frac{4}{27}P_c\left(6t+\omega^2\right)d\omega\,.
\ee
Hence, employing   Maxwell's equal area law \eqref{Maxwell}, while denoting $\omega_s$ and $\omega_l$ the `volume' of small and large black holes, we get 
the following two equations:
\ba
p&=&1+\frac{8}{3}t -\frac{8}{9}t\omega_l -\frac{4}{81}\omega_l^3=
1+\frac{8}{3}t -\frac{8}{9}t\omega_s -\frac{4}{81}\omega_s^3\,,\nonumber\\
0&=&\int_{\omega_l}^{\omega_s}\omega(6t+\omega^2)d\omega\,.\
\ea
These have a unique non-trivial solution given by $\omega_s=-\omega_l=3\sqrt{-2t}$. Hence we have found that  
\be\label{etaAdS}
\eta=V_c(\omega_l-\omega_s)=2V_c\omega_l=6V_c\sqrt{-2t}\quad \Rightarrow\quad \beta=\frac{1}{2}\,. 
\ee

To calculate the exponent $\gamma$, we differentiate Eq. \eqref{stateApV} to get 
\be
\frac{\partial V}{\partial P}\Big|_T=-\frac{9}{8}\frac{V_c}{P_c}\frac{1}{t}+O(\omega)\,.
\ee
Hence we have 
\be
\kappa_T=-\frac{1}{V}\frac{\partial V}{\partial P}\Big|_T\propto \frac{9}{8}\frac{1}{P_c}\frac{1}{t} \quad \Rightarrow \quad 
\gamma=1\,.
\ee
The critical point corresponds to the divergence of the specific heat at constant pressure
\be\label{CpCv}
C_p-C_v=-T\left(\frac{\partial P}{\partial T}\right)^2_V \left(\frac{\partial V}{\partial P}\right)_T \propto \frac{12\pi Q^2}{t}\,.
\ee
That $C_P$ but not $C_V$ diverges at our critical point can be also directly seen from the definition
\be
C_P={T}{\frac{\partial S}{\partial T}\Bigr|_P}\,,\quad C_{V}={T}{\frac{\partial S}{\partial T}\Bigr|_{V}}\,,
\ee
and the expression \cite{Dolan:2011a}
\be
T=\frac{1}{4}\frac{1}{\sqrt{\pi S}}\left(1-\frac{\pi Q^2}{S}+8PS\right)\,.
\ee
We obtain $C_{V}=0$, and 
\be
C_P=2S\frac{8PS^2+S-\pi Q^2}{8PS^2-S+3\pi Q^2}\,,
\ee
where stability requires $C_P>0$. The specific heat $C_P$ becomes singular at $8PS^2-S+3\pi Q^2=0$, and since $S=\pi r_+^2$  this occurs when 
\be
8P\pi^2r_+^4-\pi r_+^2+3\pi Q^2=0\, ,
\ee
i.e., exactly at the critical point.

Finally, the `shape of the critical isotherm' $t=0$ is given by \eqref{stateApV}
\be
p-1=-\frac{4}{81}\omega^3\quad \Rightarrow \quad \delta=3\,.
\ee 
This completes the calculation of basic critical exponents. Comparing with the previous section, where these were calculated for the Van der Waals fluid, we find that all of them coincide. 
These coefficients can also be compared with critical exponents in non-extended phase 
space, e.g., \cite{NiuEtal:2011}. {Since critical exponents may depend on the dimensionality of the system, it would be interesting to repeat our calculation for charged BTZ black hole or higher-dimensional RN-AdS black holes.}

\section{Conclusions} 
In this paper we have studied the thermodynamic behaviour of a charged AdS black hole in an extended phase space---treating the cosmological constant and its conjugate quantity as thermodynamic variables associated with the pressure and volume, respectively. For a fixed black hole charge, this identification allowed us to write the equation of state as     
$P=P(V,T)$ and study its behaviour using the standard thermodynamic techniques. 
We have shown that the system admits a first-order small--large black hole phase transition
which in many aspects resembles the liquid--gas change of phase occurring in fluids. Namely, for any non-trivial value of the charge, there exists a critical temperature $T_c$ below which this phase transition occurs. Similar to fluids, one can associate with it a coexistence line in $(P, T)$-plane which terminates at the critical point behind  which it is no longer possible to distinguish the two phases. 
We have further studied the behaviour of certain physical quantities near the critical point and calculated the corresponding critical exponents. These were shown to coincide with those of the Van der Waals fluid.

In many aspects our considerations are similar to those performed in \cite{ChamblinEtal:1999a, ChamblinEtal:1999b}; for convenience we reviewed them in the appendix. Therein the critical behaviour of charged AdS black holes was studied in the non-extended phase space and was found similar to that of the Van de Waals fluid. {However the similarity found in these papers is a mathematical analogy rather
than an exact correspondence. We point out in the appendix that one has to, for example, identify the fluid temperature with the charge of the black hole, the volume with the horizon radius and so on, confusing intensive and extensive quantities.}  On the other hand, in our approach to the problem we actually compare the same physical 
entities: the temperature of the fluid is identified with the Hawking temperature of the black hole, the volume of the fluid associates with the thermodynamic volume of the black hole etc.  Consequently the coincidence with the fluid behaviour is more direct and precise. We find a very similar behaviour of isotherms in $P-V$ plane, coexistence line in $P-T$ plane, as well the same critical exponents. All these results seem to justify and support the idea that considering the extended phase space, and hence treating the cosmological constant as a dynamical quantity, is a very interesting theoretical possibility. 

We close by noting that  we have limited ourselves to studying only the very basic properties of the simplest possible four-dimensional non-rotating charged AdS black hole.  
We have not discussed  stability issues, the calculation of fluctuations, or the construction of Landau's theory as done in \cite{ChamblinEtal:1999b}. Also, 
we have left out considering the impact of possible non-linear electrodynamics extensions of RN-AdS black hole, e.g., \cite{FernandoKrug:2003, BanerjeeRoychowdhury:2011, BanerjeeRoychowdhury:2012},   or the impact of rotation which, especially in higher dimensions where one has more rotation parameters, may make the presence of possible phase transition(s) more interesting. Also interesting would be to pursue the study \cite{NicoliniTorrieri:2011}, or to investigate the recently demonstrated isomorphism between phase structure of hairy black holes in massive gravity and RN-AdS black hole \cite{CapelaNardini:2012}. 
All these are left as future possible directions.

\vspace{0.2cm}

\section*{Acknowledgments}
This work was supported in part by the Natural Sciences and Engineering Research Council of Canada.


\appendix*



\section{Van der Waals analogy}
The thermodynamics of charged AdS black holes in the non-extended phase space was first studied in \cite{ChamblinEtal:1999a, ChamblinEtal:1999b}. The authors considered two thermodynamic ensembles, one with a fixed electric potential $\Phi$ at infinity, the other with the fixed black hole charge $Q$ at infinity. To capture the fact that $\Phi$ is an intensive quantity, whereas $Q$ is extensive, they called the corresponding ensembles as {\em grand canonical} and {\em canonical}, respectively. It turned out that the canonical ensemble admits a first order small--large black hole phase transition which is in many respects analogous to the liquid--gas phase transition in fluids. 
This triggered wide interest and various comparisons to the Van der Waals critical behaviour were studied, e.g., \cite{SahayEtal:2010a, SahayEtal:2010b, SahayEtal:2010c, NiuEtal:2011} and references therein. 
In this appendix we recapitulate this interesting analogy and relate it to the results in the extended phase space  performed in this paper.  

We shall regard the cosmological constant $\Lambda$, and hence thermodynamic pressure, as a constant and work in the ``canonical ensemble'', i.e., the action is calculated for a fixed charge $Q$. 
The analysis again starts with the equation of state \eqref{T}, which is now written as 
\be\label{beta}
\beta=\beta(r_+, Q)=\frac{4\pi r_+}{1+\frac{3r_+^2}{l^2}-\frac{Q^2}{r_+^2}}\,,
\ee
or, using the fact that the `electric potential' $\Phi$ reads $\Phi=Q/r_+$,
\be\label{PhiQ}
\beta=\beta(\Phi,Q)=\frac{4\pi Q\Phi}{l^2\Phi^2-l^2\Phi^4+3Q^2}\,.
\ee

The Euclidean action of the canonical ensemble is associated with the quantity $G$, \eqref{GibbsAdS}. To stress its dependence on the extensive quantity $Q$, we shall denote it in this appendix as $F=F(Q,\beta)$. 
This thermodynamic potential governs the behaviour of the canonical ensemble.
It reads  
\ba
F=F(Q, \beta)&=&\frac{1}{4l^2}\left(l^2r_+-r_+^3+\frac{3l^2Q^2}{r_+}\right)\,,\\
&=&\frac{1}{4l^2}\left(\frac{l^2 Q}{\Phi}-\frac{Q^3}{\Phi^3}+{3l^2Q\Phi}\right)\,,\ \ 
\ea
where $r_+$ and $\Phi$ are understood as functions of $Q$ and $\beta$, as given by Eqs. \eqref{beta} and \eqref{PhiQ}. 
Surfaces of $F$ are displayed in Fig.~10. Obviously, $F$ demonstrates the characteristic swallowtail behaviour, which corresponds to the first-order phase transition in the canonical ensemble.  [Similar behaviour, on the other hand, is absent in the grand canonical (fixed $\Phi$) ensemble.]
\begin{figure}
\begin{center}
\rotatebox{-90}{
\includegraphics[width=0.39\textwidth,height=0.34\textheight]{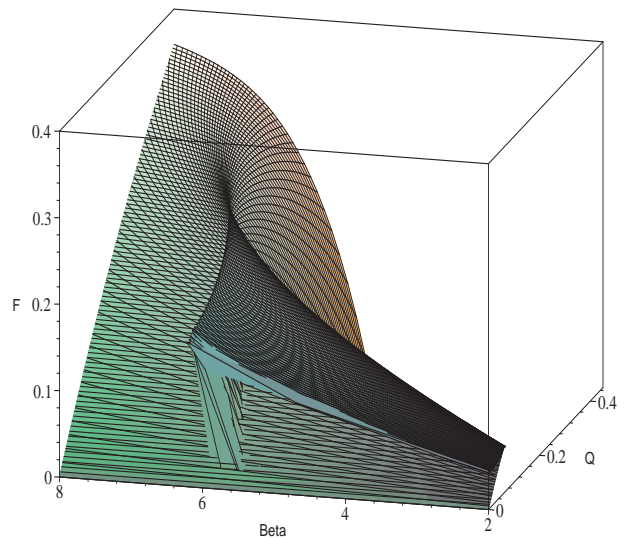}
}
\caption{{\bf Thermodynamic potential $F$.} The thermodynamic potential $F$ is depicted as a function of charge and inverse temperature for fixed $l=\sqrt{3}$.
The characteristic swallowtail behaviour corresponds to the presence of the first-order phase transition in the canonical ensemble. 
Such features are absent in the grand canonical (fixed $\Phi$) ensemble.  
}  
\end{center}
\end{figure}    

To obtain an analogy with the liquid--gas phase transition in fluids, we have to identify $F$ with the Gibbs free energy $G=G(P,T)$ of the fluid. For this we have two options, both of which were suggested in \cite{ChamblinEtal:1999a, ChamblinEtal:1999b}. One possibility is to identify $\beta$ with the pressure of the fluid and $Q$ with its temperature. We call this {\em analogy 1}:    
\be\label{A1}
\begin{array}{|c|c|}
\hline
\multicolumn{2}{|c|}{\mbox{Analogy 1}} \\
\hline
\mbox{fluid} & \mbox{AdS black hole}\\
\hline	
\mbox{temperature} & Q \\ 
\mbox{pressure} & \beta \\
\mbox{volume} & r_+\\
\hline
\end{array} 
\ee
 The other possibility is to identify $\beta$ with the fluid temperature and $Q$ with its pressure---{\em analogy 2}:
\be\label{A2}
\begin{array}{|c|c|}
\hline
\multicolumn{2}{|c|}{\mbox{Analogy 2}} \\
\hline
\mbox{fluid} & \mbox{AdS black hole}\\
\hline	
\mbox{temperature} & \beta \\ 
\mbox{pressure} & Q \\
\mbox{volume} & \Phi\\
\hline
\end{array} 
\ee
Let us study both these possibilities in greater detail.

\subsection{Analogy 1}
In this analogy, one identifies the fluid temperature with the black hole charge $Q$, and the fluid pressure with the black hole inverse temperature $\beta$. 
Since the conjugate quantity to $\beta$ is the black hole entropy $S=\pi r_+^2$, which depends entirely on the horizon radius $r_+$, it is natural to identify $r_+$ with the volume of the fluid, see table \eqref{A1}.  
The equation of state \eqref{beta}, can now be written as the following quartic for $v=r_+$:  
\be
\beta v^3-\frac{4\pi l^2}{3} v^2+\frac{\beta l^2}{3} v -\frac{l^2\beta Q^2}{3 v}=0\,,
\ee
which is going to be compared with \eqref{vdwA2}. The corresponding $\beta-r_+$ diagram is depicted on Fig.~11. This is an analogue of $p-V$ diagram of the fluid displayed in Fig.~1.
\begin{figure}
\begin{center}
\rotatebox{-90}{
\includegraphics[width=0.39\textwidth,height=0.34\textheight]{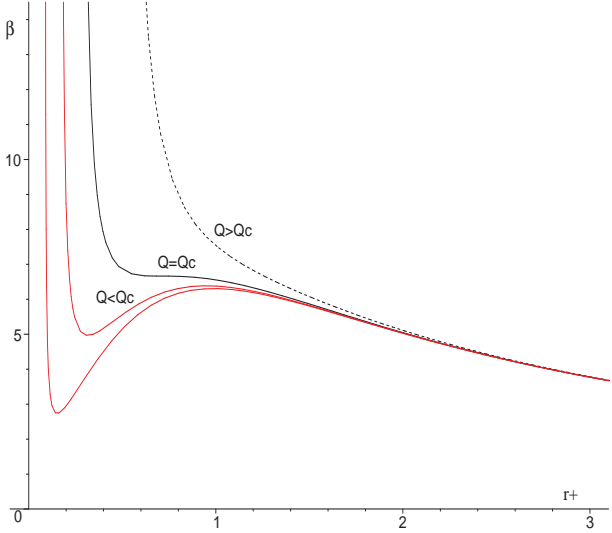}
}
\caption{{\bf $\beta-r_+$ diagram.} This is an analogue of the $p-V$ diagram of the fluid. Isocharge lines are displayed for various charges. The upper dashed line corresponds to the ``ideal gas phase'' for $Q>Q_c$, the critical isocharge line $Q=Q_c$ is highlighted by the thick solid line, lower solid lines correspond to $Q<Q_c$ for which the phase transition occurs. Cosmological constant was set $l=\sqrt{3}$.
}  
\end{center}
\end{figure}       
\begin{figure}
\begin{center}
\rotatebox{-90}{
\includegraphics[width=0.39\textwidth,height=0.34\textheight]{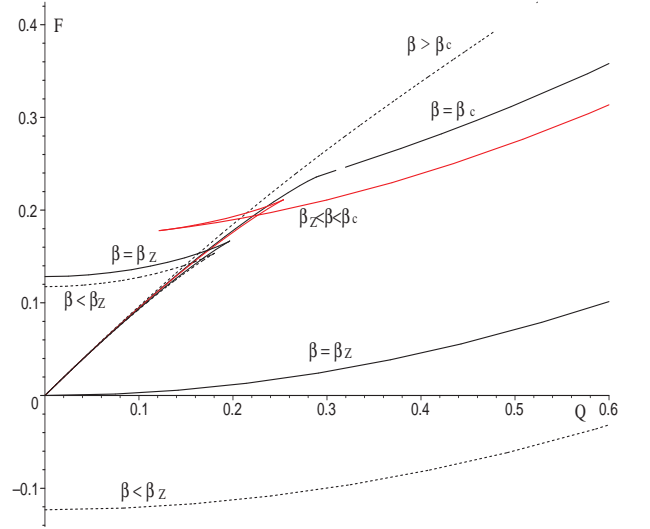}
}
\caption{{\bf ``Zorro'' diagram.} The thermodynamic potential $F$ is depicted as a function of $Q$; $l=\sqrt{3}$. Value of the inverse temperature $\beta$ decreases from top to bottom. Upper dashed line corresponds to $\beta>\beta_c$, the upper thick solid line to $\beta=\beta_c$, the lower thick solid line to ``Zorro's signature'' \cite{ChamblinEtal:1999b} at $\beta=\beta_Z$. The phase transition occurs  for $\beta_Z<\beta<\beta_c$. The bottom dashed line displays the disconnected line of  $\beta<\beta_Z$.  
}  
\end{center}
\end{figure}

The phase transition occurs for $Q<Q_c$ and $\beta_Z<\beta<\beta_c$. 
The coexistence line of the equilibrium between small and 
large black hole phase is depicted on Fig.~13. The
critical point is given by  
\be
\frac{\partial \beta}{\partial r_+}=0\,,\quad \frac{\partial^2 \beta}{\partial r_+^2}=0\,,
\ee   
which happens for 
\be
Q_c=\frac{l}{6}\,,\quad r_c=\frac{l}{\sqrt{6}}\,,\quad \beta_c=\frac{\pi l \sqrt{6}}{2}\,.
\ee
Contrary to the fluid (or extended phase space) case, there exists 
a temperature, {\em ``Zorro's'' temperature} $\beta_Z$, below which the phase transition no longer occurs. This temperature is the temperature of the 
last connected $\beta=const$ curve displayed in Fig.~12. It is given by 
\be
\beta_Z=\pi l\,.
\ee
\begin{figure}
\begin{center}
\rotatebox{-90}{
\includegraphics[width=0.39\textwidth,height=0.34\textheight]{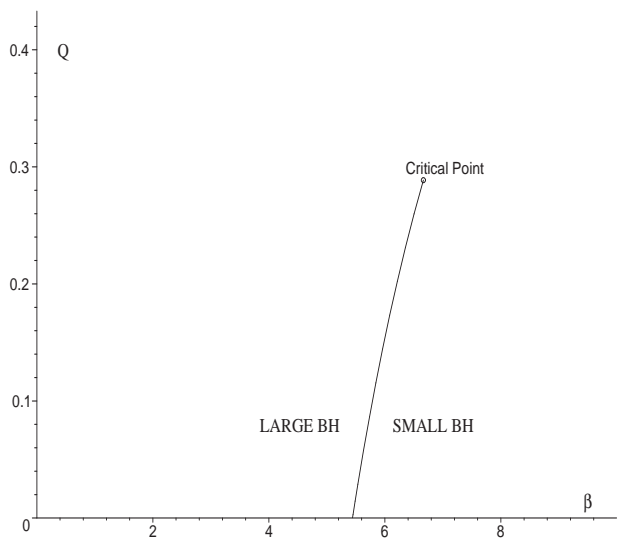}
}
\caption{{\bf Coexistence line in the $(\beta, Q)$-plane.} Here we illustrate the small-large black hole coexistence line in $(\beta,Q)$ plane for fixed $l=\sqrt{3}$, analogous to the liquid--gas coexistence line in $(p,T)$-plane.
}  
\end{center}
\end{figure}  
Another point worth  noting is the fact that the diagram $F=F(Q)$, displayed in Fig.~12, significantly differs from the corresponding fluid diagram in Fig.~4. We see that the analogy is broken here. The reason is that the swallowtail of the potential $F$, displayed on Fig.~10, `faces' the $\beta$-axis rather than the $Q$-axis. Consequently  in Fig.~13, where the coexistence line is displayed, the axes $\beta$ and $Q$ `swap' their roles to what one would expect from the corresponding fluid Fig.~5.

\subsection{Analogy 2}
\begin{figure}
\begin{center}
\rotatebox{-90}{
\includegraphics[width=0.39\textwidth,height=0.34\textheight]{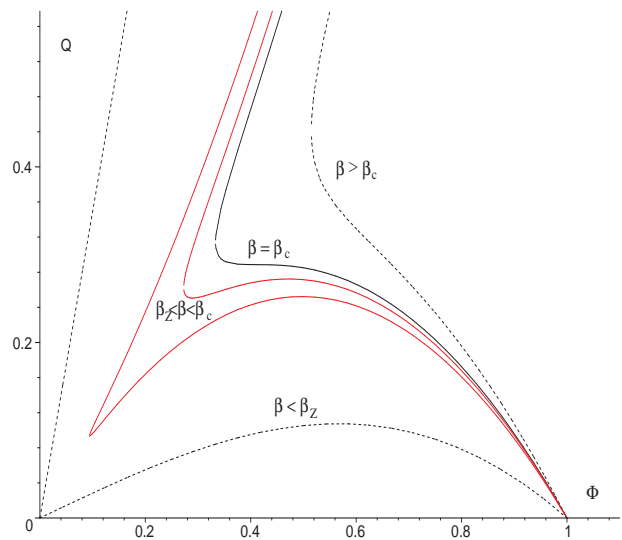}
}
\caption{{\bf $Q-\Phi$ diagram.} This is a second analogue of $p-V$ diagram of the fluid. The inverse temperature $\beta$ of isoterms decreases from top to bottom. The upper dashed line corresponds to the ``ideal gas phase'' for $\beta>\beta_c$, the critical isoterm $\beta=\beta_c$ is highlighted by the thick solid line, two lower solid lines correspond to $\beta_Z<\beta<\beta_c$ for which the phase transition occurs, the bottom dashed line displays an isoterm at $\beta<\beta_Z$. Cosmological constant was set $l=\sqrt{3}$.
}  
\end{center}
\end{figure}  
\begin{figure}
\begin{center}
\rotatebox{-90}{
\includegraphics[width=0.39\textwidth,height=0.34\textheight]{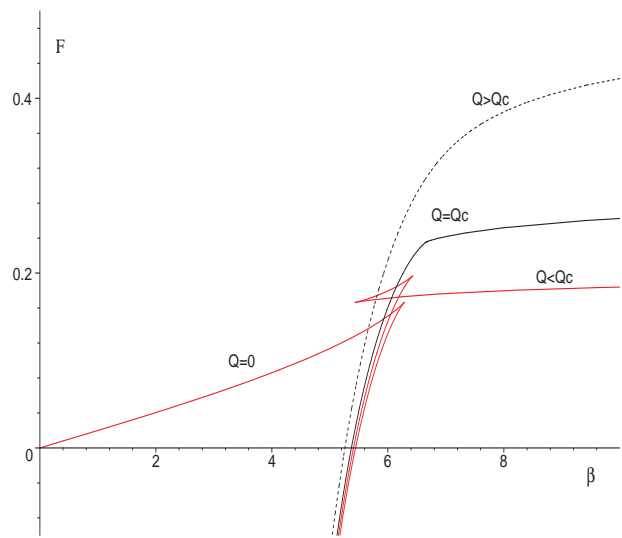}
}
\caption{{\bf Potential $F$ as function of $\beta$.} The thermodynamic potential $F$ is depicted as a function of temperature for snapshots of fixed charge and $l=\sqrt{3}$. 
The value of charge decreases from top to bottom.
Upper dashed line corresponds to $Q>Q_c$, thick solid line to $Q=Q_c$, solid line with `little bill' to $0<Q<Q_c$. The last solid line corresponds to the Schwarzschild-AdS case, $Q=0$.  The radius $r_+$ of black holes increases from right to left. In the Schwarzchild case the `branch of small black holes' disapears completely (it coincides with $\beta$-axis and represents $r_+=0$ black holes).
}  
\end{center}
\end{figure}  
In this second analogy, one identifies the fluid pressure with the black hole charge $Q$, and the fluid temperature with the black hole inverse temperature $\beta$. 
Since the conjugate quantity to $Q$ is the electric potential $\Phi$, it is reasonable to identify it with the volume of the fluid, see table \eqref{A2}.
The equation of state \eqref{PhiQ}, can now be written as the following quartic for $v=\Phi$: 
\be
Qv^3-Qv+\frac{4\pi Q^2}{\beta l^2}-\frac{3Q^3}{v l^2}=0\,,
\ee
which is to be compared with \eqref{vdwA2}. The corresponding $Q-\Phi$ diagram is depicted on Fig.~14. This is an analogue of the $P-V$ diagram in Fig.~1. 

The critical point occurs at 
\be
Q_c=\frac{l}{6}\,,\quad \Phi_c=\frac{1}{\sqrt{6}}\,,\quad \beta_c=\frac{\pi l \sqrt{6}}{2}\,.
\ee
The corresponding coexistence line is depicted again on Fig.~13. The difference is that now the axes have the ``correct position''. Also Fig.~15 is now ``quite analogous'' to Fig.~4.   

\subsection{Summary}
Both outlined analogies have many interesting features which resemble aspects of liquid--gas phase transition in the fluid system.
However, these analogies are not ``exact'': there are certain features, such as the existence of Zorro's temperature, which are not present in the fluid system. 
Moreover, the identification itself is a mathematical analogy, rather than an identification of similar physical quantities. In fact, one even has to identify quantities of `opposite' thermodynamic character: in the first analogy one identifies the extensive quantity $Q$ with the temperature of the fluid, which is an intensive quantity.
In the second analogy, one even associates intensive pressure with $Q$ and extensive volume with $\Phi$.  
These unpleasant features disappear when one uses the extended phase space analysis performed in the main text.

Let us finally mention that similar to the main text, one can study the behaviour of physical quantities near the critical point and calculate the corresponding critical exponents. This was done recently in \cite{NiuEtal:2011}, where it was found that the analogous critical exponents coincide with those of the Van der Waals fluid and are independent of the number of dimensions of the RN-AdS black hole.


\providecommand{\href}[2]{#2}\begingroup\raggedright\endgroup

\end{document}